\documentclass[12pt, a4paper, amsfonts, amssymb, amsmath, nofootinbib, showkeys, superscriptaddress]{revtex4-2}
\usepackage{color}
\usepackage{amsmath}
\usepackage{pifont}   
\usepackage{verbatim}       
\usepackage{acronym} 

\usepackage{amsmath}
\usepackage{amsfonts}
\usepackage{amssymb}
\usepackage{mathrsfs} 
\usepackage{graphicx}
\usepackage{multirow} 
\usepackage{slashed} 
\usepackage{array}
\usepackage{bbm}

\usepackage{graphicx}
\usepackage{epsf}
\usepackage{bm}
\usepackage{booktabs}
\usepackage{array}
\usepackage[caption=false]{subfig}
\usepackage[utf8]{inputenc}

\allowdisplaybreaks

\usepackage[pdftex, pdftitle={Article}, pdfauthor={Author}]{hyperref} 
\allowdisplaybreaks

\begin{document}

\title{Impact of non-thermal phase-space distributions on dark matter abundance in secluded sectors}

\author{Hugues Beauchesne}
  \email[Email address: ]{beauchesneh@phys.ncts.ntu.edu.tw}
  \affiliation{Physics Division, National Center for Theoretical Sciences, Taipei 10617, Taiwan}

\author{Cheng-Wei Chiang}
  \email[Email address: ]{chengwei@phys.ntu.edu.tw}
  \affiliation{Department of Physics and Center for Theoretical Physics, National Taiwan University, Taipei 10617, Taiwan}
  \affiliation{Physics Division, National Center for Theoretical Sciences, Taipei 10617, Taiwan}

\date{\today}

\begin{abstract}
Many new physics models include secluded sectors that interact little with the Standard Model and whose internal interactions control the dark matter abundance. If these same interactions are responsible for maintaining kinematic equilibrium within the secluded sector, it is possible that the phase-space distributions will differ considerably from their thermal values during freeze-out. This can potentially result in deviations of the dark matter abundance from that computed under the assumption of thermal distributions. In this paper, we revisit dark matter abundance computations for a benchmark secluded sector by numerically tracking the phase-space distributions. Namely, we show that the dark matter abundance can deviate considerably from standard results during the freeze-out process, but that a longer period of annihilation ultimately leaves only a slight excess.
\end{abstract}

\maketitle

\section{Introduction}\label{Sec:Intro}

The existence of dark matter (DM) is, for all intents and purposes, an established fact, but its exact nature remains unknown. Thermal freeze-out of Weakly Interacting Massive Particles (WIMP) has long been the dominant dark matter paradigm. In this mechanism, dark matter particles of masses around the electroweak scale pair annihilate to Standard Model (SM) particles until the corresponding rate drops below the expansion rate of the Universe. Unfortunately, direct detection experiments have imposed stringent constraints on WIMPs, forcing cosmologists to consider alternative scenarios. Amongst these is the possibility that dark matter is part of a larger secluded sector~\cite{Pospelov:2007mp}. In this case, dark matter can instead annihilate to new particles and the constraints from direct detection can be circumvented. Dark matter abundance is then mostly controlled by interactions that are internal to the secluded sector.

The overwhelming majority of the time, dark matter abundance computations are performed assuming the dark matter phase-space distribution is thermal, be it Maxwell-Boltzmann, Bose-Einstein or Fermi-Dirac. However, it was recently demonstrated in several papers that this assumption could break down under certain circumstances~\cite{Duch:2017nbe, Binder:2017rgn, Brummer:2019inq, Ala-Mattinen:2019mpa, Binder:2021bmg, Du:2021jcj, Ala-Mattinen:2022nuj, Hryczuk:2022gay, Chowdhury:2023jft, DiMauro:2023tho, Aboubrahim:2023yag, Bhatia:2023yux, Benincasa:2023vyp}. What happens in practice is that the processes that lead to dark matter destruction can prefer certain regions of the phase space, resulting in regions being depleted or overpopulated faster than can be balanced via equilibration processes. This typically leads to a larger amount of dark matter, as annihilation processes become less efficient, though in some cases it can lead to a smaller amount~\cite{Hryczuk:2022gay}.

The breakdown of the thermal distribution approximation is in effect caused by dark matter annihilation processes taking place faster than kinematic equilibration processes. In many secluded sectors, these processes are one and the same, or at least related. This puts into question whether previous computations of dark matter abundance in secluded sectors are justified in using the approximation of thermal distributions.

In this paper, we revisit the computation of dark matter abundances in a benchmark secluded sector by properly evolving the necessary phase-space distributions. More precisely, we study the impact of deviations from thermal phase-space distributions for codecaying dark matter.

We find the following results. During codecaying freeze-out, the phase-space distributions can differ considerably from their thermal values, leading to a substantially larger amount of dark matter at this stage. However, annihilation remains efficient for longer, leaving only a slight excess of dark matter. We therefore confirm that the use of thermal distributions is a relatively good approximation for codecaying dark matter and that the impact on previous work should be small.

The rest of this paper is organized as follows. The benchmark model is presented in Sec.~\ref{Sec:Model}. The numerical procedure is described in Sec.~\ref{Sec:NumericalProcedure}. The results are shown in Sec.~\ref{Sec:Results}. Concluding remarks are presented in Sec.~\ref{Sec:Conclusion}. A derivation of the expression for the $2 \to 2$ scattering coefficients is presented in Appendix~\ref{Sec:2to2ScatteringCoefficients}. The method used to track the integrated densities under the assumption of thermal distributions is discussed in Appendix~\ref{Sec:NumericalIntegratedDensities}.\footnote{The code is available at \href{https://github.com/HuguesBeauchesne/Non-Thermal-distributions}{https://github.com/HuguesBeauchesne/Non-Thermal-distributions}.}

\section{Model}\label{Sec:Model}

The benchmark secluded sector that we will consider is one in which the dark matter abundance is set by the so-called codecaying dark matter mechanism, which was first proposed in Ref.~\cite{Dror:2016rxc}.

Consider a set of dark matter particles and some unstable mediators that can annihilate or decay to SM particles. The mediators and dark matter particles are assumed to have similar masses. In models of codecaying dark matter, the annihilation and decay rates of the mediators to SM particles are small and the mediators decouple very early on. In contrast, the dark matter candidates and the mediators interact strongly. This results in dark matter particles and mediators maintaining chemical equilibrium even after decoupling from the SM plasma. Dark matter particles are then converted to mediators which decay back to the SM particles. The dark matter abundance is determined by when the annihilation of dark matter particles to mediators becomes inefficient.

In practice, we will concentrate on the following simple benchmark model. The dark matter candidate is a complex scalar called $\phi_A$ and is neutral under all SM gauge groups. It is kept stable by a global $U(1)$ symmetry. The mediator is a real scalar $\phi_B$ also neutral under all SM gauge groups. We assume the following Lagrangian for the new scalars
\begin{equation}\label{eq:CDMLagrangian}
  \mathcal{L} \supset-m_A^2 \phi_A^\dagger \phi_A - \frac{m_B^2}{2}\phi_B^2 -\frac{\lambda_{AB}}{2} (\phi_A^\dagger \phi_A) \phi_B^2 -
  \frac{\lambda_{Bh}}{4}\phi_B^2 h^2,
\end{equation}
where $h$ is the Higgs boson. In the next sections, we will assume dark matter of mass around the electroweak scale, which means few Higgs bosons are present in the plasma after the secluded sector has decoupled. This will ensure that little energy is exchanged between the two sectors via scatterings between Higgs bosons and $\phi_B$'s. A $\lambda_{Bh}$ term of the exact form as in Eq.~\eqref{eq:CDMLagrangian} would generally be accompanied by other interactions in a more realistic model. However, their inclusion would only affect the decoupling between the secluded and SM sectors, which could be mimicked by taking a different value of $\lambda_{Bh}$, and would not otherwise affect the much later decoupling between $\phi_A$ and $\phi_B$ in which we are interested. In addition, we will assume that $\phi_B$ decays with a decay width of $\Gamma_B$ to some particles that remain in thermal equilibrium with the SM, though the exact nature of these particles is irrelevant to the evolution equations. Because such a decay width would break a $\mathbb{Z}_2$ symmetry, it can be small in a technically natural way. Additional terms could be included in the Lagrangian, but we will ignore them for the sake of simplicity.

This benchmark is motivated by confining secluded sectors, which provide a natural setting for codecaying dark matter~\cite{Okawa:2016wrr, Beauchesne:2018myj, Beauchesne:2019ato}. For example, a confining secluded sector could contain two dark quarks that transform under a complex representation of the confining group. Chiral symmetry breaking would then result in three pseudo-Goldstone bosons, the dark pions. The neutral pion can then act as the mediator $\phi_B$ and the equivalent of the charged pion can act as the dark matter candidate $\phi_A$. In practice, the chiral Lagrangian of course contains additional terms and the couplings would be related. Furthermore, secluded sectors tend to have very universal properties, as their dynamics is generally controlled by a small subset of their lightest particles and a few parameters~\cite{Beauchesne:2019ato}. As such, even a minimal model is expected to be representative of many scenarios to a good approximation.

\section{Numerical procedure}\label{Sec:NumericalProcedure}

Consider the phase-space distribution $f_i(\vec{x}_i, \vec{p}_i, t)$ of particle $i$. This is related to the number density $n_i$ via
\begin{equation}\label{eq:nA}
  n_i = g_i \int \frac{d^3 p_i}{(2\pi)^3}f_i(\vec{x}_i, \vec{p}_i, t),
\end{equation}
where $g_i$ is the number of internal degrees of freedom of particle $i$. Under the assumptions of isotropy and uniformity, $f_i(\vec{x}_i, \vec{p}_i, t)$ is only a function of the magnitude $p_i$ of its momentum and time, i.e., $f_i(p_i, t)$.\footnote{To lighten the notation, we will generally leave implicit the $t$ in $f_i(p_i, t)$.} The evolution of $f_i$ is governed by the Boltzmann equation
\begin{equation}\label{eq:Boltzmann1}
  \left.\frac{\partial f_i}{\partial t}\right|_{p_i} - H p_i\frac{\partial f_i}{\partial p_i} = \sum \mathcal{C}[f_i(p_i)],
\end{equation}
where the time derivative is taken at constant momentum $p_i$, $\mathcal{C}[f_i(p_i)]$ are the different collision terms and $H$ is the Hubble parameter. Alternatively, the Boltzmann equation can be expressed as
\begin{equation}\label{eq:Boltzmann2}
  \left.\frac{\partial f_i}{\partial t}\right|_{p_i^c} = \sum \mathcal{C}[f_i(p_i)],
\end{equation}
where $p_i^c = a p_i$, with $a$ being the scale factor, is the comoving momentum. Both forms of the Boltzmann equation have their advantages and disadvantages. On one hand, using $p_i^c$ presents the advantage that the dark matter distribution at a given $p_i^c$ freezes at late time, which is not the case for $p_i$. On the other hand, using $p_i^c$ means that any finite grid will cover an increasingly small range of $p_i$. This can cause problems when a process produces particles at a fixed momentum, as it might eventually become outside the grid.

For the dark matter $\phi_A$, we use a grid of discrete values of comoving momentum $p_A^c$ referred to as $p^c_{A_i}$. For the mediator $\phi_B$, we use a grid of momentum $p_B$ referred to as $p_{B_i}$. Both grids are chosen to initially encompass the overwhelming majority of particles. The grid of $p_{B_i}$ is periodically updated to remove higher momenta at which the phase-space distribution has become negligible. All simulations are started at a time far before decoupling and using distributions initialized as Maxwell-Boltzmann with integrated densities and temperatures obtained using the method of Appendix~\ref{Sec:NumericalIntegratedDensities}. It was verified that the decoupling takes place at a sufficiently low temperature for the Maxwell-Boltzmann distribution to be an excellent approximation.

It is convenient to introduce a reference mass $m_{\text{ref}}$ and define $x=m_{\text{ref}}/T$, where $T$ is the temperature of the SM plasma. In practice, we will take $m_{\text{ref}}$ to be the mass of $\phi_A$. Assuming the energy content of the Universe to be dominated by the SM plasma, one can then verify that
\begin{equation}\label{eq:dtdx}
  \frac{dt}{dx} = \sqrt{\frac{45}{4\pi^3}}\frac{g_\ast^{1/2}}{h_{\text{eff}}}\frac{M_{\text{Pl}}}{m_{\text{ref}}T},
\end{equation}
where $h_{\text{eff}}$ is the effective number of entropy degrees of freedom and
\begin{equation}\label{eq:gstar}
  g_\ast^{1/2} = \frac{h_{\text{eff}}}{g^{1/2}_{\text{eff}}}\left(1 + \frac{T}{3 h_{\text{eff}}}\frac{d h_{\text{eff}}}{dT}\right),
\end{equation}
with $g_{\text{eff}}$ denoting the effective number of energy degrees of freedom. Eq.~\eqref{eq:dtdx} can be used to turn the time derivatives of Eqs.~\eqref{eq:Boltzmann1} and \eqref{eq:Boltzmann2} into the more convenient $\partial f_i/\partial x$.

In this paper, we will be interested in two types of collision terms: $1 \to 2$ decays and $2 \to 2$ scatterings. First, consider the decay
\begin{equation}\label{eq:Decay1}
  P_1 \to P_2 P_3,
\end{equation}
and its inverse. Assume $P_2$ and $P_3$ are in equilibrium. The corresponding collision term is
\begin{equation}\label{eq:Decay2}
  C[f_1(p_{1_i})] = - \frac{\Gamma_1}{\gamma_1} \left[ f_1(p_{1_i}) - f_1^{\text{eq}}(p_{1_i}) \right],
\end{equation}
where $\Gamma_1$ is the decay width of $P_1$ and $\gamma_1 = E_1/m_1$ is the usual Lorentz factor.\footnote{Consider a collision term $C[f_i]$. The convention that we adopt for the internal degrees of freedoms is that cross sections and decay widths are averaged over the internal degrees of freedom of $i$, but summed over all other internal degrees of freedom.} Second, consider the scattering
\begin{equation}\label{eq:Scattering1}
  P_1 P_2 \to P_3 P_4,
\end{equation}
and its inverse process. The collision term for $P_4$ is then
\begin{equation}\label{eq:Collisionterm2to25Copy}
  C[f_4(p_{4_i})] = \sum_{j, k} \Delta p_1 \Delta p_2 \hat{W}_{ijk} 
  \left[ f_1(p_{1_j}) f_2(p_{2_k}) - f_3(p_{3_m}) f_4(p_{4_i}) \right],
\end{equation}
where $\hat{W}_{ijk}$ are the scattering coefficients (see Appendix~\ref{Sec:2to2ScatteringCoefficients} for the precise definition and other notations). 

Several comments are in order. First, each element of $\hat{W}_{ijk}$ corresponds to a double integral and needs to be evaluated on a three-dimensional grid. It is practically impossible to perform every integral at each step. To circumvent this problem, $\hat{W}$ is evaluated on a three-dimensional grid before solving the evolution equations. During the evolution, values of $\hat{W}_{ijk}$ are evaluated using a trilinear interpolation on the grid. Second, the quantity $f_3(p_{3_m})$ that appears in the $2 \to 2$ processes is determined by fixing $p_{3_m}$ from energy conservation and linearly interpolating the $f_3$ distribution. Third, evaluating the double sum at every step can be extremely time-consuming when the grid is too fine, but using an insufficiently fine grid would lead to a large error in the computation of the number densities. We have found the best compromise between execution time and precision by taking a very fine grid and evaluating the double sum using only a subset of the points.

An adaptive step size is used as follows. An error is defined for $\phi_A$ as
\begin{equation}\label{eq:StepSizeError}
  \sigma_A = \max_i \left|\frac{f_A^{\text{RK}_2}(p_{A_i}, x + \Delta x) - f_A^{\text{RK}_1}(p_{A_i},x + \Delta x)}{f_A(p_{A_i},x)}\right|,
\end{equation}
where $f_A^{\text{RK}_1}(p_{A_i}, x + \Delta x)$ and $f_A^{\text{RK}_2}(p_{A_i}, x + \Delta x)$ are the distributions at $p_{A_i}$ evolved from $x$ to $x + \Delta x$ using the first- and (midway) second-order Runge-Kutta method, respectively. A similar error $\sigma_B$ is defined for $\phi_B$. We then define $\sigma$ as $\text{max}(\sigma_A, \sigma_B)$ if the ratio of dark matter density and entropy density is above twice its present-day value and as $\sigma_A$ otherwise. If $\sigma$ is above a predetermined tolerance~$\epsilon$, the evolution of the densities by one step is reperformed using a reduced step size of
\begin{equation}\label{eq:StepSize}
  \min\left(a \Delta x(\epsilon/\sigma)^b, c \Delta x, \Delta x^{\text{max}} \right),
\end{equation}
where we set $\epsilon = 10^{-3}$, $a = 0.9$, $b = 0.33$, $c = 1.1$ and $\Delta x^{\text{max}} = 0.1$. This is repeated until the required precision goal is met. The $f_A(p_{A_i},x)$ distribution is then updated to $f_A^{\text{RK}_2}(p_{A_i}, x + \Delta x)$ and the evolution continues using a new step size determined by Eq.~\eqref{eq:StepSize}. To determine when to end the simulation, the following quantity is introduced
\begin{equation}\label{eq:dYrel}
  \Delta Y^{\text{r}}_{A_n} = \frac{|Y_A(x_{n+1}) - Y_A(x_n)|}{Y_A(x_{n+1}) + Y_A(x_n)},
\end{equation}
where $Y_A$ is the ratio of $n_A$ and the entropy density and $x_n = x_0 + 10n$ with $x_0$ the value of $x$ at which the simulation is started and $n$ an integer. The simulation is stopped when more than 30 values of $\Delta Y^{\text{r}}_{A_n} < 10^{-3}$ have been recorded.

Finally, the validity of the numerical method has been verified by making sure that the correct dark matter abundance is reproduced in the limit where the standard integrated density approach is expected to work. This will be shown in the next section.

\section{Results}\label{Sec:Results}

\begin{figure}[htb]
\subfloat[\label{fig:YA1}]{
  \includegraphics[width=0.49\columnwidth]{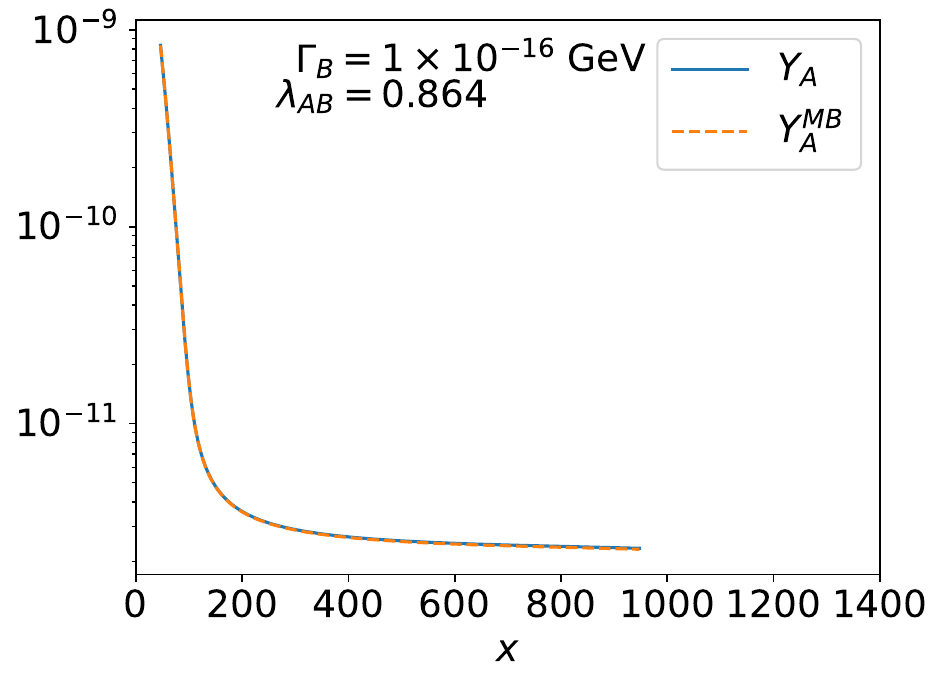}
}
\subfloat[\label{fig:YA2}]{
  \includegraphics[width=0.49\columnwidth]{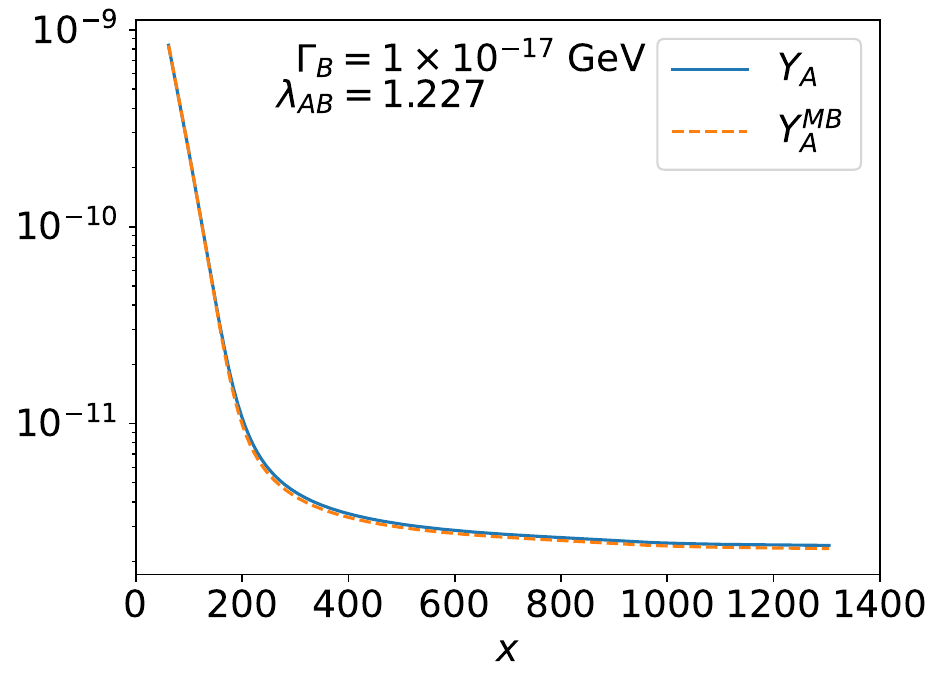}
}

\subfloat[\label{fig:YA3}]{
  \includegraphics[width=0.49\columnwidth]{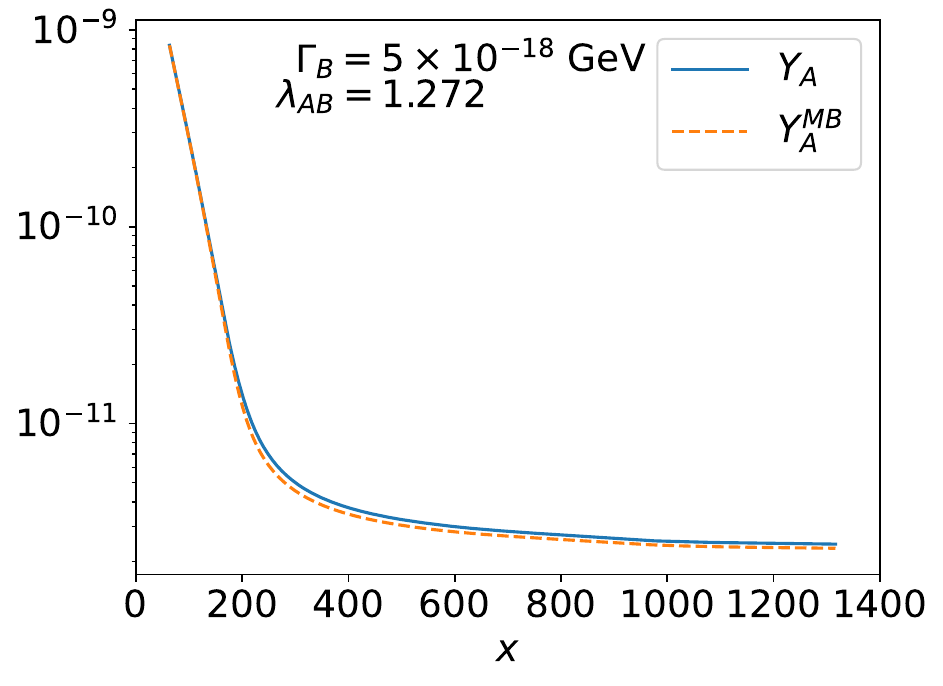}
}
\subfloat[\label{fig:YA4}]{
  \includegraphics[width=0.49\columnwidth]{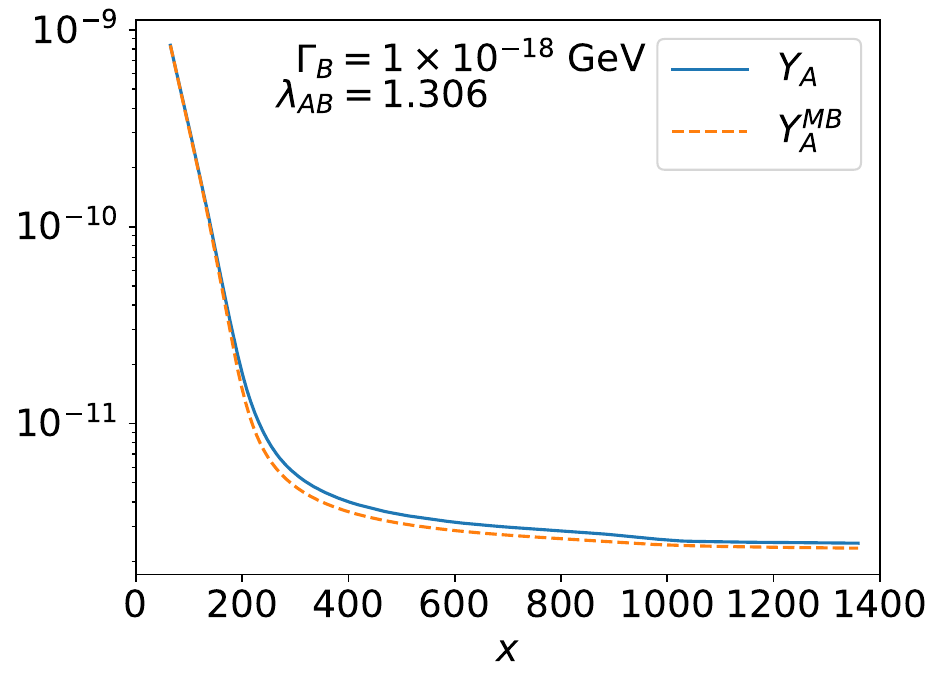}%
}
\caption{Evolution of $Y_A$ and $Y_A^{MB}$ for different decay widths $\Gamma_B$. The masses are set to $m_A = 200$~GeV and $m_B=198$~GeV. The parameter $\lambda_{Bh}$ is set to $10^{-2}$ and the parameter $\lambda_{AB}$ is adjusted for each case such that the DM abundance obtained under the assumption of thermal distributions reproduces the measured value.}\label{fig:a}
\label{fig:YA}
\end{figure}

\begin{figure}[t]
\subfloat[\label{fig:ratio}]{
  \includegraphics[width=0.49\columnwidth]{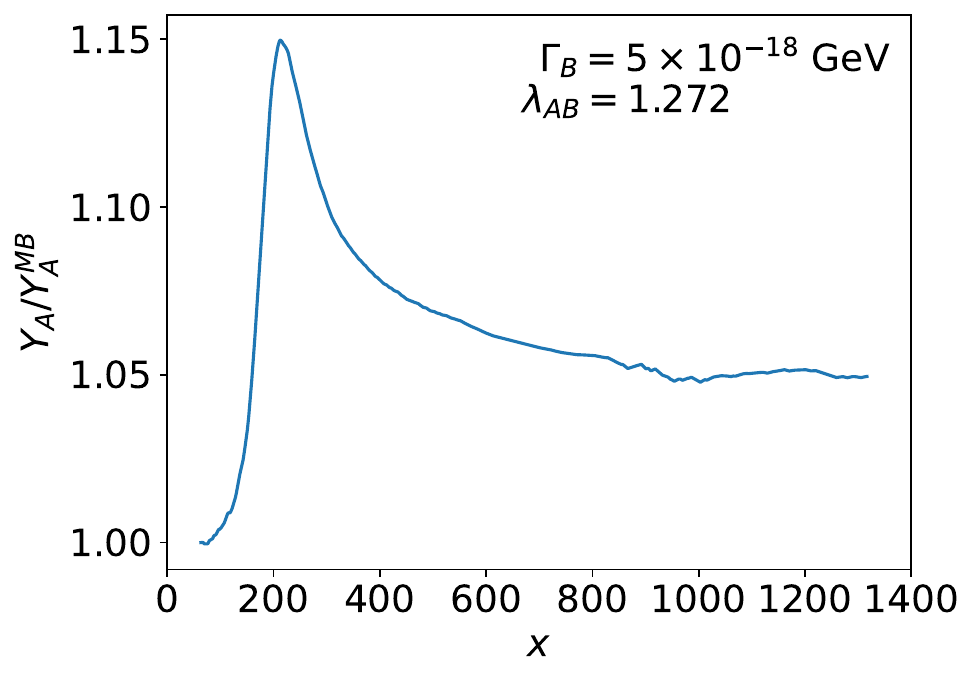}
}
\subfloat[\label{fig:fB}]{
  \includegraphics[width=0.49\columnwidth]{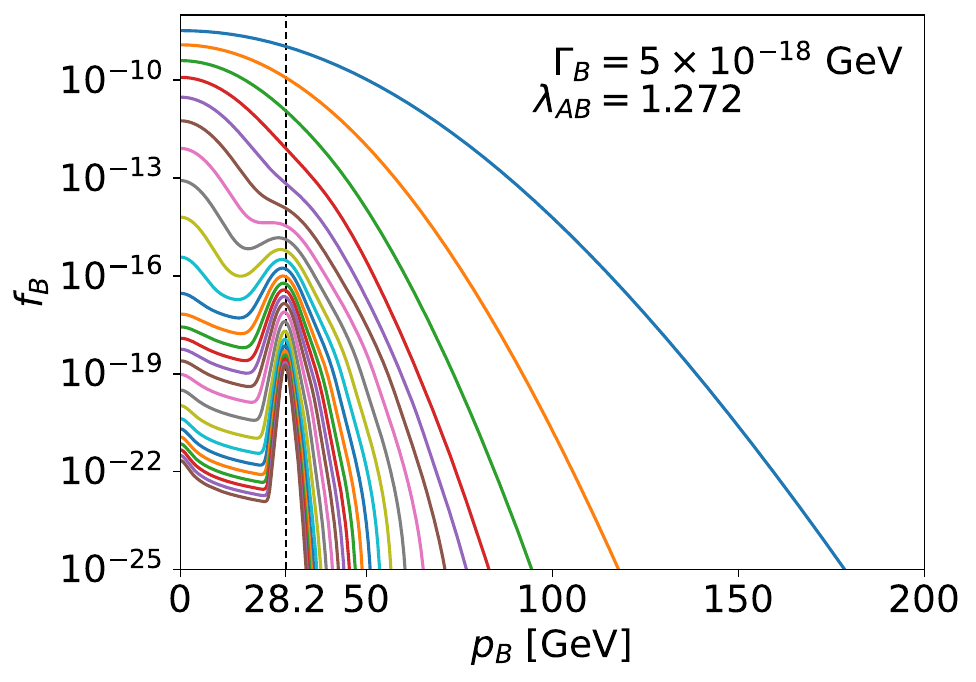}
}
\caption{(a) Example of the evolution of $Y_A/Y_A^{MB}$. (b) Example of the evolution of $f_B$. Each curve corresponds to a value of $x = x_0 + 50n$, with $x_0 \simeq 63.35$ being the $x$ at which the simulation is started for this benchmark and $n$ an integer. The topmost curve corresponds to $n=0$. Parameters are set as in Fig.~\ref{fig:YA}. The vertical line corresponds to Eq.~\eqref{eq:pPeak}.}\label{fig:others}
\end{figure}

We show in Fig.~\ref{fig:YA} different cases of the evolution of $Y_A$ and its equivalent under the approximation of Maxwell-Boltzmann distributions $Y_A^{MB}$. The dark matter $\phi_A$ is slightly heavier than the mediator $\phi_B$. The method of Appendix~\ref{Sec:NumericalIntegratedDensities} is used to compute $Y_A^{MB}$. As can be seen, the abundance begins by following very closely its thermal value. As the dark matter starts to decouple, deviations can appear and lead to a temporary excess of $\mathcal{O}(20\%)$ in some cases. However, annihilation of dark matter remains efficient for a longer period of time and the temporary excess of $Y_A$ mostly vanishes with time. Ultimately, only a small excess of dark matter is left, at best $6\%$ for our benchmarks. Fig.~\ref{fig:ratio} shows an example of the ratio $Y_A/Y_A^{MB}$. Note that both methods coincide when the decay width of $\phi_B$ becomes sufficiently large, as can be seen in Fig.~\ref{fig:YA1} and which serves as a validation of our method.

The physical reason for the deviation from the integrated density approach is best illustrated by Fig.~\ref{fig:fB}, which shows $f_B$ at different times for a benchmark point. As should be clear, the phase-space distribution of $\phi_B$ starts to deviate massively from the Maxwell-Boltzmann distribution as the system evolves. In detail, when $x$ is large, the collisions $\bar{\phi}_A \phi_A \to \phi_B \phi_B$ take place between two particles almost at rest. This leads to the production of two $\phi_B$ mediators with a peak in $f_B$ appearing at the momentum
\begin{equation}\label{eq:pPeak}
 p_{\text{peak}}\sim \sqrt{2m_B (m_A - m_B)}.
\end{equation}
The process takes place too quickly for this peak to disappear via kinetic equilibration processes. This modifies the relative rate of the $\bar{\phi}_A \phi_A \to \phi_B \phi_B$ process and its inverse, increasing the dark matter abundance. It is easy to verify that deviations start to become noticeable in Fig.~\ref{fig:YA3} around the point at which Fig.~\ref{fig:fB} deviates visibly from the Maxwell-Boltzmann distribution.

Smaller decay widths of the mediators were considered, but did not lead to much larger deviations of the dark matter abundance and proved numerically increasingly challenging. Other values of the mass splitting and $\lambda_{Bh}$ were considered, but did not lead to any qualitative differences. The small increase in the slope around $x\sim 900$ in Fig.~\ref{fig:YA} and~\ref{fig:ratio} corresponds to the QCD phase transition, where a larger amount of time passes for a given $x$ interval.

Finally, we mention that codecaying dark matter can be very difficult numerically, even under the assumption of thermal distributions. This explains the small instabilities at large $x$. Extensive numerical resources are required to obtain sufficiently accurate results, which is why few results are presented and the benchmark model is so simple.

\section{Conclusion}\label{Sec:Conclusion}

It is a distinct possibility that dark matter is part of a larger secluded sector and that the dark matter abundance is set by interactions that are internal to the secluded sector. In many such scenarios, the interactions responsible for depleting the dark matter abundance will be the same or at least related to those responsible for the kinetic equilibration of the secluded sector. It is then possible that the phase-space distributions will differ from their thermal values during freeze-out, which could affect the dark matter abundance.

In this paper, we have considered the effect of non-thermal distributions on codecaying dark matter. We find that the dark matter abundance can differ substantially from its standard abundance during the decoupling process, but that a longer period of annihilation leads to only a small increase in the final abundance. We confirmed that the use of thermal distributions is therefore a relatively good approximation for codecaying dark matter, at least as far as the final abundance is concerned.

As a concluding remark, we mention that even though we have considered only a simple example, small deviations from thermal distributions in secluded sectors should be ubiquitous. Obviously, tracking the phase-space distributions is non-trivial and might not be practical or realistic under many circumstances. At least, one should keep in mind that a proper treatment of the phase-space distributions can lead to some corrections to the dark matter abundance in many scenarios of secluded sectors. 

\section*{Acknowledgments}
This work was supported by the National Science and Technology Council under Grant No. NSTC-111-2112-M-002-018-MY3, the Ministry of Education (Higher Education Sprout Project NTU-112L104022), and the National Center for Theoretical Sciences of Taiwan.

\appendix

\section{$2 \to 2$ scattering coefficients}\label{Sec:2to2ScatteringCoefficients}

This appendix contains a computation of the $2 \to 2$ scattering coefficients that appear in Eq.~\eqref{eq:Collisionterm2to25Copy}. We combine and expand the approaches of Refs.~\cite{Du:2021jcj} and~\cite{Ala-Mattinen:2022nuj}, which are themselves based on Ref.~\cite{Hannestad:1995rs}. Consider the collision
\begin{equation}\label{eq:Collision2to2}
  P_1 P_2 \to P_3 P_4,
\end{equation}
where $P_i$ are four not necessarily distinct particles and its inverse process. The collision term for $P_4$ is given by
\begin{equation}\label{eq:Collisionterm2to21}
  C[f_4(p_4)] = \frac{S_{\text{in}}}{2E_4}\int \prod^3_{i=1} d\pi_i(2\pi)^4 \delta^4(p_1 + p_2 - p_3 - p_4)\overline{|M|^2}
  \left[ f_1(p_1) f_2(p_2) - f_3(p_3) f_4(p_4) \right],
\end{equation}
where $d\pi_i = d^3 p_i/(2E_i (2\pi)^3)$ and $S_{\text{in}}$ is a symmetry factor whose value is $1/2$ if $P_1$ and $P_2$ are identical and 1 otherwise. Consider the useful identity
\begin{equation}\label{eq:IntegralIdentity1}
  \int \frac{d^3p_3}{2E_3} = \int d^3 p_3 dE_3 \delta(E_3^2 - \vec{p_3}^2 - m_3^2)\theta(E_3).
\end{equation}
For any collision, we can define a rotated coordinate system such that
\begin{equation}\label{eq:CoordinateSystem}
  \begin{aligned}
    \vec{p}_4 &= p_4(0, 0, 1),\\
    \vec{p}_1 &= p_1(0, \sin\theta_1, \cos\theta_1),\\
    \vec{p}_2 &= p_2(\sin\beta\sin\theta_2, \cos\beta\sin\theta_2, \cos\theta_2).
  \end{aligned}
\end{equation}
Using Eqs.~\eqref{eq:IntegralIdentity1} and~\eqref{eq:CoordinateSystem}, Eq.~\eqref{eq:Collisionterm2to21} can be simplified to
\begin{equation}\label{eq:Collisionterm2to22}
  \begin{aligned}
    C[f_4(p_4)] &= \frac{S_{\text{in}}}{8(2\pi)^4 E_4}\int dp_1 d\cos\theta_1 dp_2 d\cos\theta_2 \frac{p_1^2}{E_1} \frac{p_2^2}{E_2} \theta(E_3)\\
               &  \hspace{2.0cm}\times\overline{|M|^2}
               \left[ f_1(p_1) f_2(p_2) - f_3(p_3) f_4(p_4) \right]
               \int_0^{2\pi} d\beta \delta(g),
  \end{aligned}
\end{equation}
where
\begin{equation}\label{eq:Defg}
  \begin{aligned}
    g &= E_3^2 - \vec{p_3}^2 - m_3^2\\
      &= m_1^2 + m_2^2 - m_3^2 + m_4^2 + 2E_1 E_2 - 2E_2 E_4 - 2E_1 E_4 \\
      & \hspace{0.4cm} + 2p_1 p_4 \cos\theta_1 + 2p_2 p_4 \cos\theta_2 - 2p_1 p_2(\cos\theta_1 \cos\theta_2 + \sin\theta_1 \sin\theta_2\cos\beta).
  \end{aligned}
\end{equation}
We have used the fact that the coordinate system of Eq.~\eqref{eq:CoordinateSystem} can equally have any azimuthal direction, which introduces a factor of $2\pi$. The integral on $\beta$ can easily be evaluated (including the fact that two values of $\beta$ satisfy $g(\beta)=0$), which gives
\begin{equation}\label{eq:Collisionterm2to23}
    C[f_4(p_4)] = \frac{S_{\text{in}}}{256 \pi^4 E_4 p_4^2}\int dE_1 \int dt \int dE_2 \int du \frac{\theta(E_3)\overline{|M|^2}}{\sqrt{(\partial g/ \partial \beta)^2}}
    \left[ f_1(p_1) f_2(p_2) - f_3(p_3) f_4(p_4) \right],
\end{equation}
where $\partial g/ \partial \beta$ is evaluated at
\begin{equation}\label{eq:cosbeta}
  \cos\beta^\ast = \frac{1}{2p_1 p_2 \sin\theta_1 \sin \theta_2}\left(\hat{t} + 2p_2p_4\cos\theta_2 -2p_1 p_2 \cos\theta_1\cos\theta_2\right),
\end{equation}
and we use the definitions
\begin{equation}\label{eq:Mandelstam}
  \begin{aligned}
    t       &= m_1^2 + m_4^2 - 2E_1 E_4 + 2p_1 p_4 \cos\theta_1,\\
    u       &= m_2^2 + m_4^2 - 2E_2 E_4 + 2p_2 p_4 \cos\theta_2,\\
    \hat{t} &= t + m_2^2 - m_3^2 + 2E_1E_2 - 2E_2 E_4.
  \end{aligned}
\end{equation}
The quantity $(\partial g/ \partial \beta)^2$ is given by
\begin{equation}\label{eq:dgdbeta2Special}
  (\partial g/ \partial \beta)^2 = 4p_1^2p_2^2\sin^2\theta_1\sin^2\theta_2\sin^2\beta^\ast.
\end{equation}
For computations, it can be expressed in terms of $t$ and $u$ by using Eqs.~\eqref{eq:cosbeta} and~\eqref{eq:Mandelstam}. Once a grid of discrete momenta is introduced, the collision term becomes
\begin{equation}\label{eq:Collisionterm2to24}
  C[f_4(p_{4_i})] = \sum_{j, k} \Delta E_1 \Delta E_2 W_{ijk} 
  \left[ f_1(p_{1_j}) f_2(p_{2_j}) - f_3(p_{3_m}) f_4(p_{4_k}) \right],
\end{equation}
where $m$ is known from $i$, $j$ and $k$, $W_{ijk} \equiv W(p_{4i}, p_{1j}, p_{2k})$ and
\begin{equation}\label{eq:W1}
  W(p_4, p_1, p_2) = \frac{S_{\text{in}}\theta(E_3)}{256 \pi^4 E_4 p_4^2}
  \int_{\mathcal{R}_t} dt \int_{\mathcal{R}_u} du \frac{\overline{|M|^2}}{\sqrt{(\partial g/ \partial \beta)^2}},
\end{equation}
where $\mathcal{R}_t$ and $\mathcal{R}_u$ are the domains of integration of $t$ and $u$, respectively. These intervals are defined as the region where  $|\cos\theta_i| \leq 1$ and $(\partial g/ \partial \beta)^2 \geq 0$. The latter condition is equivalent to requiring the existence of a physical (i.e., real) $\beta^\ast$ angle for the given values of $\theta_1$ and $\theta_2$. Note that $(\partial g/ \partial \beta)^2$ can be expressed as
\begin{equation}\label{eq:dgdbetaabs2}
  (\partial g/ \partial \beta)^2 = a\cos^2\theta_2 + b \cos\theta_2 + c,
\end{equation}
where
\begin{equation}\label{eq:abc}
    a = 4p_2^2(t - (E_1 - E_4)^2),\quad
    b = -4p_2\hat{t}(p_4 - p_1 \cos\theta_1),\quad
    c = 4p_1^2 p_2^2 \sin^2\theta_1 - \hat{t}^2.
\end{equation}
Since $a = -4p_2^2|\vec{p}_1 - \vec{p}_4|^2 \leq 0$, $(\partial g/ \partial \beta)^2$ can only be non-negative when
\begin{equation}\label{eq:Delta1}
  \Delta = b^2 - 4 a c
\end{equation}
is greater or equal to 0. This quantity can be rewritten as
\begin{equation}\label{eq:Delta2}
  \Delta = -16p_1^2 p_2^2 \sin^2\theta_1(\hat{a} t^2 + \hat{b}t + \hat{c}),
\end{equation} 
where
\begin{equation}\label{eq:abchat}
    \hat{a} = 1, \quad
    \hat{b} = 4E_2 E_3 - 2(m_2^2 + m_3^2), \quad
    \hat{c} = 4(E_2 - E_3)(E_2 m_3^2 - E_3 m_2^2) + (m_2^2 - m_3^2)^2. 
\end{equation}
Define 
\begin{equation}\label{eq:Delta3}
  \hat{\Delta} = \hat{b}^2 - 4 \hat{a} \hat{c}.
\end{equation} 
The quantity $\hat{\Delta}$ must be positive for $\mathcal{R}_t$ and $\mathcal{R}_u$ to be non-trivial. Conveniently,
\begin{equation}\label{eq:Delta4}
  \hat{\Delta} = 16 p_2^2 p_3^2.
\end{equation}
Therefore,  satisfying $\hat{\Delta} > 0$ and $E_3 > 0$ is equivalent to satisfying $E_3 > m_3$. Considering the results above, the interval $\mathcal{R}_t$ is given by
\begin{equation}\label{eq:Rt}
  \mathcal{R}_t =
    \begin{cases}
      \emptyset & \text{if } \hat{\Delta} \leq 0, \\
      [t_{\text{min}}^1, t_{\text{max}}^1] \cap  [t_{\text{min}}^2, t_{\text{max}}^2] & \text{otherwise},
    \end{cases}       
\end{equation}
where
\begin{equation}\label{eq:Rt2}
  \begin{aligned}
  t_{\text{max$\slash$min}}^1 &=  m_1^2 + m_4^2 - 2 E_1 E_4 \pm 2 p_1 p_4, \\
  t_{\text{max$\slash$min}}^2 &= m_2^2 + m_3^2 - 2 E_2 E_3 \pm 2 p_2 p_3.
  \end{aligned}
\end{equation}
This result is expected, as it is simply a rewording of the condition that the angle between $P_1$ and $P_4$ and the angle between $P_2$ and $P_3$ must both be real. The interval $\mathcal{R}_u$ can be simplified by noting that $(\partial g/ \partial \beta)^2>0$ automatically implies the other requirement $-1<\cos\theta_2 < 1$ inside $\mathcal{R}_t$. This can be seen from Eq.~\eqref{eq:dgdbeta2Special}. Having both $\cos^2\theta_2 > 1$ and $(\partial g/ \partial \beta)^2>0$ would require both $\sin\theta_2$ and $\sin\beta^\ast$ to be purely imaginary. This requirement is incompatible with Eq.~\eqref{eq:cosbeta}, as the left-hand side would be real and the right-hand side purely imaginary. Then, $\mathcal{R}_u$ is given by
\begin{equation}\label{eq:Ru}
  \mathcal{R}_u =
      [u_{\text{min}}, u_{\text{max}}]
\end{equation}
where
\begin{equation}\label{eq:Ru2}
    u_{\text{max$\slash$min}} = m_2^2 + m_4^2 - 2 E_2 E_4 + 2 p_2 p_4 x_{1/2}, \quad
    x_{1/2}                      = \frac{-b \mp \sqrt{\Delta}}{2a}.
\end{equation}
Finally, we can rewrite Eq.~\eqref{eq:Collisionterm2to24} in terms of momenta as
\begin{equation}\label{eq:Collisionterm2to25}
  C[f_4(p_{4_i})] = \sum_{j, k} \Delta p_1 \Delta p_2\hat{W}_{ijk} 
  \left[ f_1(p_{1_j}) f_2(p_{2_k}) - f_3(p_{3_m}) f_4(p_{4_i}) \right],
\end{equation}
where
\begin{equation}\label{eq:What}
  \hat{W}_{ijk} = \frac{p_1}{E_1} \frac{p_2}{E_2} W_{ijk}.
\end{equation}

\section{Evolution equations with thermal distributions}\label{Sec:NumericalIntegratedDensities}

When the phase-space distributions are assumed to be thermal, the dark matter abundance can be calculated by using the formalism of Ref.~\cite{Beauchesne:2021opx}. This paper provides analytical expressions for annihilation and energy exchange rates in general 2-to-2 scatterings under the assumption of Maxwell-Boltzmann distributions and does not assume that the incoming particles share the same temperature. In this section, we summarize the important results of this formalism and express them in the notation of this paper.

Assume a 2-to-2 process
\begin{equation}\label{eq:2to2process}
 P_1 P_2 \to P_3 P_4,
\end{equation}
with differential cross section $d\sigma /dt$ where $t$ is the standard Mandelstam variable.\footnote{Ref.~\cite{Beauchesne:2021opx} uses $t=(p_1 - p_3)^2$ and we maintain this convention in this section. Keep in mind that $t=(p_1 - p_4)^2$ was used in Sec.~\ref{Sec:2to2ScatteringCoefficients} to be consistent with Ref.~\cite{Du:2021jcj} on which that section was based.} Define the energies
\begin{equation}\label{eq:Energyvariables}
    E_+ = E_1 + E_2, \quad E_- = E_1 - E_2, \quad E_+' = E_3 + E_4, \quad E_-' = E_3 - E_4,
\end{equation}
the center-of-mass momenta
\begin{equation}\label{eq:pij}
  \begin{aligned}
    p_{12} &= \frac{\sqrt{(s - (m_1 + m_2)^2)(s - (m_1 - m_2)^2)}}{2\sqrt{s}},\\
    p_{34} &= \frac{\sqrt{(s - (m_3 + m_4)^2)(s - (m_3 - m_4)^2)}}{2\sqrt{s}},
  \end{aligned}
\end{equation}
where $\sqrt{s}$ is the center-of-mass energy and the temperatures
\begin{equation}\label{eq:TSTA}
  T_S = \frac{2T_1 T_2}{T_2 + T_1}, \;\;\; T_A = \frac{2T_1 T_2}{T_2 - T_1},
\end{equation}
where $T_1$ and $T_2$ are respectively the temperatures of $P_1$ and $P_2$. All necessary rates are then given by
\begin{align}
    \langle\sigma v\rangle_{P_1 P_2 \to P_3 P_4}^{T_1, T_2} &= \frac{T_A \int_{s_{\text{min}}}^\infty ds \int_{\sqrt{s}}^\infty dE_+\left[e^{-A_+} - e^{-A_-}\right]p_{12}\sqrt{s}\sigma(s) }{8 T_1 T_2 m_1^2 m_2^2 K_2\left(\frac{m_1}{T_1}\right) K_2\left(\frac{m_2}{T_2}\right)},\nonumber\\
    \langle\sigma v E_+\rangle_{P_1 P_2 \to P_3 P_4}^{T_1, T_2} &= \frac{T_A \int_{s_{\text{min}}}^\infty ds \int_{\sqrt{s}}^\infty dE_+ E_+\left[e^{-A_+} - e^{-A_-}\right]p_{12}\sqrt{s}\sigma(s) }{8 T_1 T_2 m_1^2 m_2^2 K_2\left(\frac{m_1}{T_1}\right) K_2\left(\frac{m_2}{T_2}\right)},\nonumber\\
    \langle\sigma v E_-\rangle_{P_1 P_2 \to P_3 P_4}^{T_1, T_2} &= \frac{T_A \int_{s_{\text{min}}}^\infty ds \int_{\sqrt{s}}^\infty dE_+\left[B_+ e^{-A_+} - B_- e^{-A_-}\right]p_{12}\sqrt{s}\sigma(s) }{8 T_1 T_2 m_1^2 m_2^2 K_2\left(\frac{m_1}{T_1}\right) K_2\left(\frac{m_2}{T_2}\right)},\nonumber\\
    \langle\sigma v E_+'\rangle_{P_1 P_2 \to P_3 P_4}^{T_1, T_2} &= \frac{T_A \int_{s_{\text{min}}}^\infty ds \int_{\sqrt{s}}^\infty dE_+ E_+\left[e^{-A_+} - e^{-A_-}\right]p_{12}\sqrt{s}\sigma(s) }{8 T_1 T_2 m_1^2 m_2^2 K_2\left(\frac{m_1}{T_1}\right) K_2\left(\frac{m_2}{T_2}\right)},\label{eq:ThermalAv2T}\\
    \langle\sigma v E_-'\rangle_{P_1 P_2 \to P_3 P_4}^{T_1, T_2} &= \frac{T_A \int_{s_{\text{min}}}^\infty ds \int_{\sqrt{s}}^\infty dE_+ \frac{(m_3^2 - m_4^2)}{s}E_+\left[e^{-A_+} - e^{-A_-}\right]p_{12}\sqrt{s}\sigma(s) }{8 T_1 T_2 m_1^2 m_2^2 K_2\left(\frac{m_1}{T_1}\right) K_2\left(\frac{m_2}{T_2}\right)}\nonumber\\
   &\hspace{0.4cm} +\frac{T_A \int_{s_{\text{min}}}^\infty ds \int_{\sqrt{s}}^\infty dE_+\left[C_+ e^{-A_+} - C_- e^{-A_-}\right]p_{34}\sqrt{s}\sigma^t(s) }{8 T_1 T_2 m_1^2 m_2^2 K_2\left(\frac{m_1}{T_1}\right) K_2\left(\frac{m_2}{T_2}\right)},\nonumber
\end{align}
where $v$ is the M{\o}ller velocity, $K_n(x)$ are the modified Bessel functions of the second kind, $s_{\text{min}}=\text{max}((m_1 + m_2)^2, (m_3 + m_4)^2)$,
\begin{equation}\label{eq:ABC}
  \begin{aligned}
    &A_\pm = \frac{E_+}{T_S} + \frac{E_-^{\text{min/max}}}{T_A},&
    B_\pm &= T_A + E_-^{\text{min/max}},\\
    &C_\pm = T_A \mp 2p_{12}\sqrt{\frac{E_+^2 - s}{s}}, &
    E_-^{\text{min/max}} &= \frac{E_+(m_1^2 - m_2^2)}{s} \mp 2p_{12}\sqrt{\frac{E_+^2 - s}{s}},
  \end{aligned}
\end{equation}
and
\begin{equation}\label{eq:FAM}
  \sigma^t = \int_{t_1}^{t_0} \frac{d\sigma}{dt}\cos\theta_3 dt = \int_{t_1}^{t_0}\frac{d\sigma}{dt}\left[1 + \frac{t - t_0}{2p_{12}p_{34}} \right] dt,
\end{equation}
with
\begin{equation}\label{eq:t0t1}
  t_0(t_1) = \left[\frac{m_1^2 - m_2^2 - m_3^2 + m_4^2}{2\sqrt{s}}\right]^2 - \left(p_{12} \mp p_{34}\right)^2.
\end{equation}
Using the rates from Eq.~\eqref{eq:ThermalAv2T}, it is a simple task to write and solve the Boltzmann and Friedmann equations for the integrated number densities $n$ and energy densities $\rho$. The temperatures are computed by using the fact that $\rho/n$ is independent of the chemical potential and the results
\begin{equation}\label{eq:nrho}
    n^\text{eq}(T)    = \frac{gm^2T  }{2\pi^2}K_2(m/T),\quad
    \rho^\text{eq}(T) = \frac{gm^2T  }{2\pi^2}\left[ mK_1(m/T) + 3TK_2(m/T) \right].
\end{equation}

\bibliography{biblio}
\bibliographystyle{utphys}

\end{document}